\documentstyle[12pt,aps]{revtex}

\begin{document}
\draft\onecolumn
\title{Polynomial procedure\\ of avoiding multiqubit errors\\  
arising due to qubit-qubit interaction}
\author{L. Fedichkin}
%{L. Fedichkin}
\address{Institute of Physics and Technology\\
Russian Academy of Sciences\\
Nakhimovsky pr. 34, Moscow, 117218, Russia\\
E-mail: leonid@ftian.oivta.ru}

\date{Received \today}

\maketitle

\begin{abstract}
{Recently proposed implementations of quantum computer suffer from
unavoidable interaction between quantum bits depending upon
data being written in them. Novel procedure of
avoiding multiqubit errors
arising due to uncontrollable qubit-qubit interaction 
by using addititonal intermediate qubits is proposed.
It is shown that the scheme requires only polynomial
increase in number of qubits and algorithmic steps.}
\end{abstract}

Since the pioneering works of Yu.~Manin~\cite{Manin} and
R.~Feynman~\cite{Feynman82} there has been a tremendous progress
in quantum computation theory. P.~Shor~\cite{Shor} and
L.~Grover~\cite{Grover} have discovered fast quantum algorithms
of great practical importance.
One of the most important reasons why
the experimental realization of practically useful large scale
quantum computer is not attained until now is
attitude of quantum computers (compared to classical ones)
to various types of errors.
Hopefully there are known quantum error correction
procedures~\cite{e1,e2,e3,e4,e5,e7,e8,Vatan}
which help to correct errors which occurred simultaneously in single
quantum bit (qubit) or in few qubits due to interaction with environment or
imprecise implementation of local gates.

        In 1998 J.~Gea-Banacloche~\cite{Gea} revealed
the significance of another source of errors:
internal interaction in quantum computer between neighbour qubits.
The interaction serves to entangle qubits if necessary,
so it should be switched on every time while implementing
two-qubit gate and be switched off otherwise. But the accuracy of switcher
amplitude could not be generally higher than several orders of magnitude.
We should also control moments of switching off/on with such accuracy.
Therefore there is weak unavoidable interaction between qubits any time.
J.~Gea-Banacloche~\cite{Gea} noted that even such
a weak interaction leads to errors which could completely destroy
quantum computer performance in large scale quantum registers.
These errors differ from common few-qubit errors since
they originate from internal qubit-qubit interaction rather than
from influence of noisy environment. They conserve coherence
of quantum computer but
spreads over whole quantum register and make
initially unentangled blocks of qubits to entangle each other.
In 1999 it was pointed out by the same author~\cite{Gea99}
that common error correction methods do not solve the problem
since these procedures imply that the probability of
whole quantum register to be entangled due to errors during
time of performing one of basic gates
(I will denote this time as $\tau$)
is negligible.
In the subsequent discussion it is demonstrated how to avoid these errors
if we have managed to build scalable quantum computer.

As it was shown in~\cite{Gea99} the interaction between
neighbour qubits $i$ and $j$ frequently leads to
Hamiltonian which in the basis
$\left\{\left| 00 \right\rangle,\left| 01 \right\rangle,
\left| 10 \right\rangle,\left| 11 \right\rangle\right\}$
%($\left| 0 \right\rangle$ and $\left| 1 \right\rangle$ states
%correspond to opposite orientations of dipoles)
will have a matrix like the following
\begin{equation}
H_{ij} =
 \left( \begin{array}{cccc}   
  a^2& 0& 0&  0\\
    0&ab& 0&  0\\
    0& 0&ab&  0\\
    0& 0& 0&b^2\\
 \end{array} \right)   .
\end{equation}
The inequality $a\ne b$ results in nonadditive interaction energy.
It is convenient to split interaction Hamiltonian into
additive $H_A$ and nonadditive parts $H_N$, where
\begin{equation}
 H_A =
 \left( \begin{array}{cccc}   
  a^2& 0& 0&  0\\
    0&\frac{a^2+b^2}{2}& 0&  0\\
    0& 0&\frac{a^2+b^2}{2}&  0\\
    0& 0& 0&b^2\\
 \end{array} \right)  ,
\end{equation}
\begin{equation}
 H_N = \hbar\delta \omega
 \left( \begin{array}{cccc}   
    0&  0& 0&  0\\
    0&  1& 0&  0\\
    0&  0& 1&  0\\
    0&  0& 0& 0\\
 \end{array} \right) 
\end{equation}
where
\begin{equation}
 \delta \omega = \frac{(a-b)^2}{2\hbar}.
\end{equation}
The dimensionless parameter $\delta$ ($\delta = \tau \delta \omega$)
can be used  to evaluate entanglement during one computing step.
The action of $H_A$ does not entangle qubits.
Moreover, by going to an interaction  picture with
state $\left| 0 \right\rangle$ having additional energy $a^2/2$
and state $\left| 1 \right\rangle$ having additional energy $b^2/2$
additive part can be effectively removed.
The numerical value of factor $\delta \omega$ in $H_N$
depends on qubits being used. In some cases non-additive part can contain
also  off-diagonal terms, whose influence on
computation performance are similar to diagonal terms~\cite{Gea99}.
The property  of all such terms which is of  importance for further discussion
is their decrease with the increase of distance between interacting qubits $R$
according to:
\begin{equation}  \label{R3}
 \delta = O \left(R^{-3}\right),\qquad {at}\quad{large}\quad R.
\end{equation}
This property results from
dipole-dipole interaction energy scaling and it is valid
if qubits interact at large distances like dipoles or weaker.
It is mostly the case. For qubits with stronger
entangling interaction such as proposed ones in recent paper~\cite{Udrea}
(among numerous qubit proposals
I did not find other examples  of such kind), where it was proposed to
choose for $\left| 0 \right\rangle$ state the absense
of electron in semiconductor quantum wire
and to choose for $\left| 1 \right\rangle$ state the presense of electron,
Eq.~(\ref{R3}) is not true if qubits interaction is not screened by electrodes
and direct implementation of technique given below does not help.
But proposal of Ionicioiu {\em  et al.}~\cite{Udrea} can be reduced to
dipole case by doubling (i.e. by polynomial increase) number of qubits. 
The reduction is implemented by  encoding quantum information into pairs of qubits
instead of single ones
\[
 \alpha \left| 0 \right\rangle + \beta \left| 1 \right\rangle
  \longmapsto
  \alpha \left| 01 \right\rangle + \beta \left| 10 \right\rangle
\]
to provide presense of qubit and its ``antiqubit'' in each qubits pair.

        The degradation of computer perfornmance due to errors
is characterized by probability to get right outcome after
measurement --- quality factor $Q$:
\begin{equation}
 Q = \sum_{m \in S}\left| \left\langle m|\Psi \right\rangle \right|^2
\end{equation}
where $S$ is set of problem solutions,
$\Psi$ is wavefunction of quantum register before
final measurement. In abovementioned papers~\cite{Gea,Gea99}
it is shown that quality degrades as
\begin{equation}
 Q \propto \exp\left(-\sigma^2\right)
\end{equation}
where dispersion $\sigma$ is given by
\begin{equation}
 \sigma = C P\sqrt{L} \,\delta  .
\end{equation}
Here $C$ is some constant
whose exact value depends on algorithm being implemented and
input data (usually of order of unity~\cite{Gea,Gea99}), $P$ is number of algorithm steps, $L$ is number of qubits.
The uncontrollable qubit-qubit interaction is usually weak, i.e.
$\delta \ll 1$, but
the numerical factor $P\sqrt{L}$ is large.
For example, to factor number of 1000 decimal digits
(using nowadays in RSA public key cryptography procedure)
by implementing Shor algorithm~\cite{Shor} $L$ should be about $10^4$ and
$P$  should exceed $5\times10^6$, therefore $P\sqrt{L}$ would exceed
$5\times10^8$. In this evaluation
we do not include any error correction procedures
which will also increase with both $P$ and $L$ in polynomial way.
It should be also noted that at reasonable quantum register sizes $L$
$\bigl($at $L > 1\left/\delta^2\right.\bigr)$ computer strongly degrades
during even single computational step. It makes impossible application
of any error correction procedures to improve computing quality.

The scalability of quantum computer means that we can assemble
as large uniform quantum register as we need,
but can not significantly change interaction between qubits.
Although obtained results can be easily generalized to
two-dimensional and three-dimensional layouts of qubits,
to be more concrete we restrict our consideration to one-dimensional
case when all qubits are located along straight line forming
one-dimensional grid with constant  distance $r$ between neighbours.

To avoid errors of interaction
it is proposed to substitute logical ideal 
(non-interacting when it is not needed)
qubits by sets of interacting qubits in the following way.
Each logical qubit with number $k$ and value $a_k$
$\left| a_k \right\rangle$ is encoded by
$m$ real qubits
\begin{equation}
 \left| a_k \right\rangle \longmapsto
  | a_k \overbrace{00\ldots 0}^{m-1}\,\rangle .
\end{equation}
The logical quantum register $\left| a_1 a_2 \ldots a_L \right\rangle$ 
is encoded then 
by $mL$ real qubits as following: 
\begin{equation}
 \left| a_1 a_2 \ldots a_L \right\rangle \longmapsto
  | a_1 \overbrace{00\ldots 0}^{m-1}a_2 \overbrace{00\ldots 0}^{m-1}
    \ldots a_L \overbrace{00\ldots 0}^{m-1}
  \,\rangle .
\end{equation} 
All one-qubit logical gates $V_k$ are performed as usual 
\begin{equation}
 V_k \longmapsto V_{(k-1)m+1}
\end{equation}
taking into account shift $k \longmapsto (k-1)m+1$ of qubits numbers. 
Additionally in order to avoid external errors 
all intermediate qubits in state $\left| 0 \right\rangle$ 
should be measured during each computational step. 
As they are not entangled to others we  can measure them 
without disturbing quantum coherent state of register. 
Nontrivial logical two-qubit gate $W_{k,k+1}$ between neighbours 
is performed via sequence of basic swap operators $S_{n,n+1}$ 
\[ 
 S_{n,n+1} \left| a_n a_{n+1} \right\rangle = \left| a_{n+1} a_n \right\rangle
\]
\begin{equation}
 S_{(k-1)m+1,km} = 
  \overbrace{S_{km-1,km}\, S_{km-2,km-1} \,
    \ldots S_{(k-1)m+1,(k-1)m+2}}^{m-1}  
\end{equation}
and one nontrivial two-qubit gate $U_{km,km+1}$.
\begin{equation}
 W_{k,k+1} \longmapsto S_{(k-1)m+1,km}\, U_{km,km+1} \, S_{(k-1)m+1,km}.  
\end{equation}
So it is performed via $2m-1$ basic gates. 
Since all qubits in state $\left| 0 \right\rangle$ are not in superposition 
state,  
they are prohibited from interaction entanglement~\cite{Gea}. 
By introducing such procedure  
interaction between neighbour qubits  
at distance $r$ is effectively replaced by interaction between 
qubits at distance $mr$. 
Two qubits in superpositon states approach each other 
only during implementation two-qubit gates but 
in this case additional known entanglement is not errorneous and 
can lead only to slight (and known in advance)  
change of  nontrivial two-qubit gate.  

Consider the influence of  proposed procedure 
on quantum computer performance upon parameter $m$. 

Needed space resources $L^\prime$ (number  of qubits) are linear increased: 
\[ L^\prime=mL. \]
Time resources $P^\prime$ (number of basic gates) are also 
linear increased: 
\[ P^\prime \le (2m-1)P.\] The equality is attained when only two-qubit 
gates are applied. The effective qubit-qubit interaction constant 
is decreased according to Eq.~[\ref{R3}] 
\[
 \delta^\prime \le \delta \left/m^3 \right.
\]
Therefore dispersion of computation quality $\sigma^\prime$ is 
polynomially improved 
\begin{equation}
 \sigma^\prime \le \sigma m^{-3/2}
\end{equation}
So by polynomial (upon needed dispersion change) 
increase of parameter $m$ computation quality can be improved 
at any given qubit-qubit interaction. 
Finally, novel error avoiding 
procedure is proposed. It allows to  
operate with qubits interacting each 
other by polynomial increase of space and time resources.

\end{document}